%% file: xmr_btc_atomic_swaps.tex
\documentclass[runningheads]{llncs}
\input{packages}
\input{macros}

\begin{document}
\title{Atomic Swaps between Bitcoin and Monero}
%
%
\author{Philipp Hoenisch\inst{1,2} \and
Lucas Soriano del Pino\inst{1,2}}
\authorrunning{Hoenisch et al.}
%
\institute{COMIT,
\email{\{firstname\}@comit.network} \and
CoBloX Pty Ltd, 
\email{\{firstname\}@coblox.tech}
}
\maketitle              
\begin{abstract}

Due to the evergrowing blockchain ecosystem, interoperability has become a matter of great importance. 
Atomic swaps allow connecting otherwise isolated blockchains while adhering to the core principles of censorship resistance and permissionlessnes. 
Up until recently, atomic swap protocols have mostly relied on complex script support, excluding certain types of blockchains.
With advances in cryptography, it is now possible to build a bridge between almost any two blockchains. 
In this work, we give an explanation of one such protocol which applies adaptor signatures on Bitcoin to procure atomic swaps between Monero and Bitcoin. 
We dive into the cryptographic details, discuss its limitations and give an outlook on our current work where we use adaptor signatures on the Monero signature scheme. 

\end{abstract}
\input{introduction}

\input{htlc_atomic_swaps}

\input{old_protocol}

\input{new_protocol}

\input{conclusion}

%
%

\bibliographystyle{splncs04}
\bibliography{mybibliography}
\end{document}

%% file: packages.tex
\usepackage[utf8]{inputenc}
\usepackage[T1]{fontenc}
\usepackage{amsmath}
\usepackage{mdframed}
\usepackage{tikz}
\usetikzlibrary{patterns,shapes,arrows,fit,backgrounds,positioning,shadows}
\usepackage[n, advantage, operators, sets,
adversary, landau, probability, notions, logic, ff, mm, primitives,
events, complexity, asymptotics, keys]{cryptocode}

\usepackage{xcolor}
\usepackage[capitalize]{cleveref}

\usepackage{dashbox}
\usepackage{pgf-umlsd}
\usepackage{xurl}

%% file: macros.tex
\newcommand{\checkRelative}[1]{+#1}

\newcommand{\minipagesize}{2.2cm}

\tikzstyle{MSsingle} = [rectangle, draw,  
text width=6.5em, text centered, rounded corners]

\tikzstyle{TX} = [rectangle, draw, text width=5em, text centered, rounded corners, minimum height=4.5em, minimum width=6em]
\tikzstyle{TXempty} = [rectangle, text width=5em, text centered, rounded corners, minimum height=4em, minimum width=6em]
\tikzstyle{TXemptysmall} = [rectangle, text width=5em, text centered, rounded corners, minimum height=1.5em, minimum width=1.5em]
\tikzstyle{TXsmall} = [rectangle, draw, text centered, rounded corners, minimum height=2em]

\tikzstyle{MO} = [circle, draw]

\tikzstyle{IF} = [diamond, draw, fill=gray!20]
\tikzstyle{output} = [rectangle, draw]

\tikzstyle{party} = [rectangle, draw, fill=gray!20]

\tikzstyle{line} = [draw, -latex']

\tikzstyle{lables} = [midway, above, sloped]

\tikzstyle{labelAbove} = [above, pos=0.8]
\tikzstyle{labelAbove_small} = [above, pos=0.75, font=\small]

\tikzstyle{labelBelow} = [below, pos=0.8]
\tikzstyle{labelBelow_small} = [below, pos=0.75, font=\small]

\usepackage[misc,clock,geometry]{ifsym}

\newcommand{\alice}{\textbf{Alice}}
\newcommand{\bob}{\textbf{Bob}}
\newcommand{\wallet}[1]{\textsf{wallet}_#1}
\newcommand{\bitcoin}{\textsf{bitcoin}}

\newcommand{\fundrawtransaction}[1]{\textsf{fundrawtransaction}(#1)}

\newcommand{\Pdleq}[3]{\pcalgostyle{P}_{\textsf{DLEQ}(#1, #2, #3)}}
\newcommand{\Vdleq}[3]{\pcalgostyle{V}_{\textsf{DLEQ}(#1, #2, #3)}}

\newcommand{\btc}{\textsf{amt}_{\textsf{btc}}}
\newcommand{\xmr}{\textsf{amt}_{\textsf{xmr}}}
\newcommand{\lock}[1]{tx_{\textsf{lock}}^{\textsf{#1}}}
\newcommand{\txoutput}[2]{txout_{\textsf{#1}}^{\textsf{#2}}}
\newcommand{\redeem}{tx_{\textsf{redeem}}^{\textsf{btc}}}
\newcommand{\refund}{tx_{\textsf{refund}}^{\textsf{btc}}}
\newcommand{\cancel}{tx_{\textsf{cancel}}^{\textsf{btc}}}
\newcommand{\punish}{tx_{\textsf{punish}}^{\textsf{btc}}}
\newcommand{\Spend}[2]{S_{#1}^{#2}}
\newcommand{\View}[1]{V_{#1}}
\newcommand{\view}[1]{v_{#1}}
\newcommand{\s}[1]{s_{#1}}
\newcommand{\signature}[2]{\sigma_{\textsf{#1}}^{#2}}
\newcommand{\encsig}[2]{\hat{\sigma}_{\textsf{#1}}^{#2}}
\newcommand{\expiry}[1]{t_{#1}}
\newcommand{\address}[2]{\textsf{addr}_{\textsf{#1}}^{\textsf{#2}}}
\newcommand{\A}{A}
\newcommand{\B}{B}
\newcommand{\dleq}[1]{\pi_{#1}}
\newcommand{\buildtx}{\textsf{Tx}}
\newcommand{\buildtxout}{\textsf{TxOut}}
\newcommand{\ecdsa}{\textsf{ECDSA}}
\newcommand{\pcreturnspace}{\phantom{\pcreturn (}}
\newcommand{\fee}{\textsf{fee}}
\newcommand{\rec}{\textsf{Rec}}
\renewcommand{\verify}{\textsf{Vrfy}}

%% file: introduction.tex
\section{Introduction}

Since the birth of Bitcoin in 2008\cite{nakamoto2008}, many other cryptocurrencies have been introduced. 
It is without a doubt that this flourishing ecosystem has evolved into an enormous financial market. 
Cryptocurrencies are traded against fiat (e.g. USD, AUD, EUR) or against each other. 
However, due to the lack of interoperability between different blockchains, most of the trades are executed on centralized exchanges. 
Due to regulations, these centralized exchanges have integrated complex KYC (Know Your Customer) procedures where traders have to go through lengthy processes to prove their identity. 
In addition, traders give up control over their hard-earned coins by depositing them in the exchange so that they can execute trades.
The trader now has to trust the exchange to manage their funds according to the highest standards, to protect them against thieves or not lose them otherwise. 
This trust was misused more than once in the past and billions of dollars in user funds have been lost\cite{exchangeHacks}. 

One could say that these centralized exchanges are now a relic of the past. 
A new era of decentralized exchanges has started, adhering to the core idea of Bitcoin: censorship resistance at all levels. 

Decentralized exchanges powered by atomic swaps, first introduced in 2015 by TierNolan\cite{TierNolan2013}, can now promise more guarantees in terms of security and privacy to traders. 

The original idea of atomic swaps uses HTLCs (Hash Time-Lock Contracts), imposing certain requirements on the underlying blockchains: (1) they must support scripts so that one can build hash locks; and (2) they must support timelocks. 

Technology has evolved and, with advances in cryptography, a new way of cross-chain atomic swaps using adaptor signatures is gaining traction.  

Atomic swaps using adaptor signatures (also referred to as Scriptless Scripts) have several advantages over traditional atomic swaps using HTLCs: 
(1) contrary to HTLCs where the same hash has to be used on each chain, transactions involved in an atomic swap using adaptor signatures cannot be linked; and 
(2) since no script is involved, the on-chain footprint is reduced which makes the atomic swap cheaper.

Within this work we present our current efforts on cross-chain atomic swaps using adaptor signatures. 
In particular, we show how adaptor signatures can be employed to swap between Monero and Bitcoin. Notably, the former does not support scripts or timelocks. 

%% file: htlc_atomic_swaps.tex
\section{HTLC-based Atomic Swaps}
Replacing centralized exchanges by decentralized ones is not new. 
The idea of using HTLCs for atomically swapping two assets across two chains has been around for a while\cite{TierNolan2013}. Various companies have used this technology in their products and protocols for cross-chain trading\cite{comit,opendex}. Moreover, HTLCs are also used in the Lightning Network for multi-hop payments\cite{lightning2016}. 

In a nutshell, an HTLC-based atomic swap protocol works like this: we assume two parties, Alice and Bob found each other somehow and agreed on the amounts and assets (e.g. bitcoin and ether) which the two parties want to exchange. Alice generates a random secret $s$ and uses a cryptographic hash function to generate hash $h$. She then creates an HTLC using $h$ and locks up the bitcoin. These coins can either be redeemed (spent) using the secret $s$ or are returned to her after time $t$ has passed. Bob does the same thing on the other chain: he locks up his ether in an HTLC using the same hash $h$.

Since Alice knows the original secret $s$ that was used to produce the hash $h$, she can redeem the ether from Bob's HTLC. By doing so, she reveals the secret $s$ to Bob who can then take the bitcoin, completes the swap.

This apparently simple process has a few drawbacks: 
\begin{itemize}
    \item The requirements on the underlying blockchains are high. A certain script capability is required in order to support a hash function as well as timelocks. While many blockchains support these two features, some lack either one (e.g. Grin has no script support and hence no support for hash functions) or both (e.g. Monero). 
    \item By definition, the same hash has to be used on both chains. This allows an independent third party to link those two transactions. Worse yet, since blockchain transactions are publicly available to everyone, this onlooker can now track where the parties move their newly acquired funds. 
    \item The use of scripts (e.g. on Bitcoin, Litecoin, etc) or smart contracts (e.g. on Ethereum) results in an increased on-chain footprint and higher transaction fees in general.
\end{itemize}

With recent advancements in cryptography and the application of adaptor signatures to atomic swaps, it is now possible to overcome almost all of the aforementioned drawbacks. 
For example, Grin-Bitcoin swaps can be realized despite Grin's lack of a scripting language. Using Schnorr adaptor signatures and timelocks, an atomic swap protocol can be executed\cite{comit_grin_btc}. 

Recently, Gugger, J. (aka \textit{h4sh3d}) came up with a protocol which enables atomic swaps between Monero and Bitcoin\cite{gugger2020}. In the next section we discuss this protocol in detail; in \cref{new_protocol}, we present our current work, motivated by some of the limitations of \cite{gugger2020}.

%% file: old_protocol.tex
\section{BTC to XMR atomic swaps}\label{old_protocol}

\input{schemas/old_tranasction_schem}

The protocol described in this section is largely based on the work of Gugger\cite{gugger2020}.
We highlight key differences between the original and our instantiation of it\cite{xmr-btc-comit} throughout.

\subsection{Situation}
Alice and Bob have agreed to a trade in which Alice will send $\xmr$ to Bob, and Bob will send $\btc$ to Alice.
They require this exchange to be atomic, i.e. the change of ownership of one asset should effectively imply the change of ownership of the other.
Additionally, should the exchange not come to fruition, they expect any committed assets to be returned to them.

\subsection{Overview}

\subsubsection{Happy path}
After exchanging a set of addresses, keys, zero-knowledge proofs and signatures, Bob locks up $\btc$ in a Point Time Locked Contract (PTLC)\cite{PTLC} locked using point $\Spend{a}{btc}$ by publishing $\lock{btc}$.
Being a PTLC, the output is also spendable in an alternative manner after time $\expiry{1}$.


Alice subsequently locks up $\xmr$ in a shared output with public spend key $\Spend{a}{xmr} + \Spend{b}{xmr}$ and public view key $\View{a}$ + $\View{b}$ by publishing $\lock{xmr}$.
The relationship between $\Spend{a}{xmr}$ and $\Spend{a}{btc}$ is that they share the same secret key $\s{a}$ despite being points on different elliptic curve groups.
The same relationship applies to $\Spend{b}{xmr}$ and $\Spend{b}{btc}$.
This output will be owned by the party with knowledge of both $\s{a}$ and $\s{b}$.

Bob notices the publication of $\lock{xmr}$ and sends to Alice an adaptor signature\cite{one-time-ves} $\encsig{redeem}{\Spend{a}{btc},B}$ which she is able to combine with $\s{a}$ producing $\signature{redeem}{B}$.
Provided there is enough time until $\expiry{1}$, she then publishes a $\redeem$ signed with $\signature{redeem}{B}$ and her own $\signature{redeem}{A}$.
Broadcasting this transaction moves $\btc$ to an address owned by Alice.

Finally, Bob sees $\redeem$ on the blockchain.
He finds $\signature{redeem}{B}$ in the witness stack and combines it with $\encsig{redeem}{\Spend{a}{btc},B}$ to learn $\s{a}$.
With knowledge of both $\s{a}$ and $\s{b}$, Bob is the de facto owner of $\xmr$, which he is able to move to a different address of his at any time.

\subsubsection{Cancel}

Once Bob has published $\lock{btc}$, if time $\expiry{1}$ is reached, either party can elect to publish $\cancel$, diverging from the ``happy path''.
The transaction $\cancel$ was constructed in such a way that it will only be mined after time $\expiry{1}$.
The use of transaction-level timelocks is one of the ways in which this protocol deviates from the original\cite{gugger2020}.

\paragraph{Refund:}

With $\cancel$ confirmed on the blockchain, Bob should immediately publish $\refund$ to reclaim his $\btc$ (minus some fees).

Alice would then spot $\refund$ on the Bitcoin blockchain, giving her access to $\signature{refund}{A}$.
Combining $\signature{refund}{A}$ with $\encsig{refund}{\Spend{b}{btc},A}$ would leak $\s{b}$ to her.
Knowing $\s{a}$ and $\s{b}$, Alice would effectively reclaim control over $\xmr$, which she could eventually move back to one of her wallet addressess.

\paragraph{Punish:}

Should Bob remain inactive after $\cancel$ is published, Alice still has a way to get compensation for the failed swap.
After time $\expiry{2}$, Alice can punish Bob for not triggering the refund path in time by publishing $\punish$.
With this transaction Alice claims $\btc$.
The $\xmr$ remains locked forever, but from Alice's perspective it is as if the trade went through.

The existence of $\punish$ therefore incentivises Bob to publish $\refund$ as soon as possible.
Either way, Alice, the party who has no agency on whether refund will occur or not, remains protected.

\subsection{Off-chain preparation}\label{preparation}

As hinted at in the previous section, before Alice and Bob can go on-chain they must exchange some data.

For simplicity we assume a fixed $\fee$ for all Bitcoin transactions that must be signed by both parties.
In practice, the best way to handle transaction fees would be to adopt a Child-pays-for-parent (CPFP)\cite{cpfp} strategy, so that the parties do not have to commit to a particular fee rate ahead of time.

\subsubsection{Key generation}\label{sec:btc-key-gen}


Firstly, they engage in a key generation protocol, as shown in \cref{fig:key_gen_protocol}.

Alice sends to Bob a Bitcoin public key $\A$; a Monero private view key $\view{a}$; a Monero public spend key $\Spend{a}{xmr}$; a Bitcoin public key $\Spend{a}{btc}$; and a Discrete Logarithm Equality (DLEQ) $\dleq{\s{a}}$ proof between $\Spend{a}{xmr}$ and $\Spend{a}{btc}$.
The characteristics of this kind of proof will be explained in greater detail in \cref{DLEQ}.

Similarly, Bob sends to Alice a Bitcoin public key $\B$; a Monero private view key $\view{b}$; a Monero public spend key $\Spend{b}{xmr}$; a Bitcoin public key $\Spend{b}{btc}$; and a DLEQ proof $\dleq{\s{b}}$ between $\Spend{b}{xmr}$ and $\Spend{b}{btc}$.

If either party receives an invalid DLEQ proof, they must abort the protocol.

\begin{figure}[ht]
  \centering
  \begin{mdframed}
    \begin{center}
      \scalebox{0.8}{\procedure{$\Pi_{\kgen}$}{
        \alice \< \< \bob \\
        a \sample \ZZ_q; \A \gets a G \< \< \\
        \view{a} \sample \ZZ_p \\
        \s{a} \sample \ZZ_p \\
        \Spend{a}{btc} \gets \s{a} G; \Spend{a}{xmr} \gets \s{a} H \\
        \dleq{\s{a}} \gets \Pdleq{(G, \Spend{a}{btc})}{(H, \Spend{a}{xmr})}{\s{a}} \\
        \< \sendmessageright*{(\A, \view{a}, \Spend{a}{btc}, \Spend{a}{xmr}, \dleq{\s{a}})} \< \\
        \< \< \Vdleq{(G, \Spend{a}{btc})}{(H, \Spend{a}{xmr})}{\dleq{\s{a}}} \stackrel{?}{=} 1 \\
        \< \< b \sample \ZZ_q; \B \gets b G \< \< \\
        \< \< \view{b} \sample \ZZ_p \\
        \< \< \s{b} \sample \ZZ_p \\ 
        \< \< \Spend{b}{btc} \gets \s{b} G; \Spend{b}{xmr} \gets \s{b} H \\
        \< \< \dleq{\s{b}} \gets \Pdleq{(G, \Spend{b}{btc})}{(H, \Spend{b}{xmr})}{\s{b}} \\
        \< \sendmessageleft*{(\B, \view{b}, \Spend{b}{xmr}, \Spend{b}{btc}, \dleq{\s{b}})} \< \\
        \Vdleq{(G, \Spend{b}{btc})}{(H, \Spend{b}{xmr})}{\dleq{\s{b}}} \stackrel{?}{=} 1 \\
        \pcreturn (a, \A, \B, \view{a}, \view{b}, \s{a}) \< \< \pcreturn (b, \A, \B, \view{a}, \view{b}, \s{b}) \\
      }}
    \end{center}
  \end{mdframed}
  \caption{Key generation protocol.}
  \label{fig:key_gen_protocol}
\end{figure}

\subsubsection{Address exchange}

Additionally, Alice sends to Bob two Bitcoin addresses $\address{redeem}{A}$ and $\address{punish}{A}$; and Bob sends to Alice one Bitcoin address $\address{refund}{B}$.
The $\btc$ will end up in one of these depending on the protocol execution.

\subsubsection{Expiries}

The value of the two timelocks $\expiry{1}$ and $\expiry{2}$ must be confirmed before Alice and Bob can sign any transactions.
Timelock $\expiry{1}$ determines how long Alice will have to publish and confirm $\lock{xmr}$, and safely redeem $\lock{btc}$.
Timelock $\expiry{2}$ determines how long Bob has to refund his bitcoin after $\cancel$ is published by either party.

In this protocol we only use relative timelocks because they create consistent windows of action no matter when $\lock{btc}$ and $\cancel$ are included in a block. 

\subsubsection{Signing phase}\label{sec:btc-signing-phase}

This phase is a pre-requisite to Bob being able to lock up the bitcoin safely.
It also ensures that Alice can safely lock up the monero herself, once she has confirmed that the bitcoin is on the blockchain.

Before either party can start signing the Bitcoin transactions, Bob must define what $\lock{btc}$ looks like.
Given what they both already know, they can construct the PTLC output: one which can be spent by providing signatures for $\A$ and $\B$.
Bob builds the rest of the transaction using a Bitcoin wallet which will contribute the necessary inputs and outputs.
This is the first step of the signing protocol, which is depicted in \cref{fig:signing_protocol}.
Bob sends the unsigned $\lock{btc}$ to Alice, alongside the signatures $\signature{cancel}{B}$ and $\signature{punish}{B}$.
He can safely share these signatures with Alice because $\lock{btc}$ remains unpublished and unsigned.

With $\lock{btc}$ Alice computes the signatures $\signature{cancel}{A}$ and $\signature{punish}{A}$.
She also computes the adaptor signature $\encsig{refund}{\Spend{b}{btc},A}$, which Bob would need to decrypt if he ever wants to refund his bitcoin.
Using the corresponding decrypted signature $\signature{refund}{A}$ to publish $\refund$ would leak $\s{b}$ to Alice, allowing her to refund her own monero.
Alice sends back $\signature{cancel}{A}$ and $\encsig{refund}{\Spend{b}{btc},A}$ to Bob.

All that remains is for Bob to compute his own $\signature{cancel}{B}$.

\begin{figure}[tp]
  \centering
  \begin{mdframed}
    \begin{center}
      \scalebox{0.55}{\procedure{$\Pi_{\sig}(\A, \B, \btc, \expiry{1}, \expiry{2}, \fee)$}{
        \alice(a, \Spend{b}{btc}) \< \< \bob(b) \\
        \< \< \pccomment{Generating the Bitcoin lock transaction} \\
        \< \< \txoutput{lock}{btc} \gets \buildtxout(\A + \B, \btc) \\
        \< \< \lock{btc} \gets \begin{subprocedure} \dbox{
            \procedure{${\wallet{\bitcoin}.\tiny\fundrawtransaction{\txoutput{\tiny lock}{\tiny btc}}}$}{
              {\tiny tx \gets \fundrawtransaction{\txoutput{\tiny lock}{\tiny btc}}} \\
              \tiny\pcreturn \tiny tx}}
        \end{subprocedure} \\
        \< \< \pccomment{Signing the cancel transaction} \\
        \< \< \txoutput{cancel}{btc} \gets \buildtxout(\A + \B, \btc - \fee) \\
        \< \< \cancel \gets \buildtx(\lock{btc}, \txoutput{cancel}{btc}, \expiry{1}) \\
        \< \< \signature{cancel}{B} \gets \ecdsa.\sig(b, \cancel) \\
        \< \< \pccomment{Signing the punish transaction} \\
        \< \< \txoutput{punish}{btc} \gets \buildtxout(\address{punish}{A}, \btc - 2 \cdot \fee) \\
        \< \< \punish \gets \buildtx(\lock{btc}, \txoutput{punish}{btc}, \expiry{2}) \\
        \< \< \signature{punish}{B} \gets \ecdsa.\sig(b, \punish) \\
        \< \sendmessageleft*{\lock{btc}, \signature{cancel}{B}, \signature{punish}{B}} \< \\
        \pccomment{Signing the cancel transaction} \\
        \txoutput{cancel}{btc} \gets \buildtxout(\A + \B, \btc - \fee) \\
        \cancel \gets \buildtx(\lock{btc}, \txoutput{cancel}{btc}, \expiry{1}) \\
        \signature{cancel}{A} \gets \ecdsa.\sig(a, \cancel) \\
        \pccomment{Generating adaptor signature for refund transaction} \\
        \txoutput{refund}{btc} \gets \buildtxout(\address{refund}{B}, \btc - 2 \cdot \fee) \\
        \refund \gets \buildtx(\cancel, \txoutput{refund}{btc}, \cdot) \\
        \encsig{refund}{\Spend{b}{btc},A} \gets \ecdsa.\enc\sig(a, \Spend{b}{btc}, \refund) \\
        \pccomment{Signing the punish transaction} \\
        \txoutput{punish}{btc} \gets \buildtxout(\address{punish}{A}, \btc - 2 \cdot \fee) \\
        \punish \gets \buildtx(\lock{btc}, \txoutput{punish}{btc}, \expiry{2}) \\
        \signature{punish}{A} \gets \ecdsa.\sig(a, \punish) \\
        \< \sendmessageright*{\signature{cancel}{A}, \encsig{refund}{\Spend{b}{btc},A}} \< \\
        \< \< \pccomment{Signing the refund transaction} \\
        \< \< \txoutput{refund}{btc} \gets \buildtxout(\address{refund}{B}, \btc - 2 \cdot \fee) \\
        \< \< \refund \gets \buildtx(\cancel, \txoutput{refund}{btc}, \cdot) \\
        \< \< \signature{refund}{B} \gets \ecdsa.\sig(b, \refund) \\
        \pcreturn ((\cancel, \signature{cancel}{A}, \signature{cancel}{B}), \< \< \pcreturn ((\cancel, \signature{cancel}{A}, \signature{cancel}{B}), \\
        \pcreturnspace (\punish, \signature{punish}{A}, \signature{punish}{B})) \< \< \pcreturnspace (\refund, \encsig{refund}{\Spend{b}{btc},A}, \signature{refund}{B}), \\
        \< \< \pcreturnspace \lock{btc})
      }}
    \end{center}
  \end{mdframed}
  \caption{Signing protocol. Both parties must verify the signatures received, but this is left out for clarity.}
  \label{fig:signing_protocol}
\end{figure}

\subsection{On-chain protocol}\label{sec:on-chain-protocol}

In \cref{preparation} we have explained how Alice and Bob set the stage for the swap to take place.
The sequence diagram in \cref{fig:onchain_seq} shows the rest of the steps towards a successful atomic swap.

With the ability to broadcast signed versions of $\cancel$ and $\refund$ to take his coins back, Bob can now proceed by publishing $\lock{btc}$.
He uses his Bitcoin wallet to sign each input and broadcasts it to the network.

Alice finds $\lock{btc}$ on the blockchain by using the transaction ID which can be deterministically computed from $\lock{bitcoin}$.
With enough confirmations on it to consider it irreversible and sufficient time until $\expiry{1}$, Alice publishes $\lock{xmr}$.
The only requirement on this transaction is that it must pay $\xmr$ to the address corresponding to the public spend key $\Spend{a}{xmr} + \Spend{b}{xmr}$ and the public view key $\View{a} + \View{b}$.
Bob does not need to know any other details, because the parties do not need to sign transactions depending on $\lock{xmr}$ ahead of time.

Bob finds $\lock{xmr}$ on the blockchain by leveraging his knowledge of the private view key $\view{a} + \view{b}$.
In Monero, only parties with knowledge of the private view key are privy to transactions involving the matching address. Once Bob considers that $\lock{xmr}$ has garnered enough confirmations, he proceeds by sending $\encsig{redeem}{\Spend{a}{btc},B}$ to Alice.
This adaptor signature can be decrypted by Alice to grant her the ability to redeem $\btc$.

On receiving $\encsig{redeem}{\Spend{a}{btc},B}$ Alice first verifies that what she has received is useful to her by executing  $\ecdsa.\enc\verify(\B, \Spend{a}{btc}, \redeem, \encsig{redeem}{\Spend{a}{btc},B})$.
This ensures that the adaptor signature commits to a valid signature on $\B$ for the transaction $\redeem$, encrypted by $\Spend{a}{btc}$.
With knowledge of $\s{a}$ Alice decrypts it by calling $\ecdsa.\dec(\s{a}, \encsig{redeem}{\Spend{a}{btc},B})$, obtaining $\signature{redeem}{B}$.

Alice now has the means to publish $\redeem$, but she must only do so if there is enough time to confirm the transaction before $\expiry{1}$.
Otherwise, Bob could front-run her transaction with $\cancel$, ensuring his refund of $\btc$ and still finding $\redeem$ in the mempool, with which he would be able to also claim the $\xmr$.
Assuming there is enough time, she goes ahead and publishes $\redeem$, claiming $\btc$.

Finally, Bob can use the information obtained by the publication of $\redeem$ to claim the $\xmr$.
He takes the transaction from the blockchain, extracts the signature $\signature{redeem}{B}$ from it and calls $\ecdsa.\rec(\signature{redeem}{B}, \encsig{redeem}{\Spend{a}{btc},B})$ to obtain $\s{a}$.
As the sole owner of both $\s{a}$ and $\s{b}$, Bob is the only one capable of moving $\xmr$ to a different address.
He does so at his own convenience, so that he can safely forget $\s{a} + \s{b}$.

\begin{figure}[ht]
  \centering
  \begin{mdframed}
    \begin{center}
      \scalebox{0.65}{\begin{sequencediagram}
        \newinst{xmr}{Monero}
        \newinst[4]{a}{Alice}
        \newinst[2]{b}{Bob}
        \newinst[4]{btc}{Bitcoin}

        \mess{b}{$\lock{btc}$}{btc}
        \begin{call}
          {a}
          {look for $\lock{btc}$}
          {btc}{$\lock{btc}$}
        \end{call}
        \mess{a}{$\lock{xmr}$}{xmr}
        \begin{call}
          {b}
          {look for $\lock{xmr}$}
          {xmr}{$\lock{xmr}$}
        \end{call}
        \postlevel
        \mess{b}{$\encsig{redeem}{\Spend{a}{btc},B}$}{a}
        \postlevel
        \begin{call}
          {a}
          {$\ecdsa.\dec\sig(\s{a}, \encsig{redeem}{\Spend{a}{btc},B})$}
          {a}
          {$\signature{redeem}{B}$}
        \end{call}
        \mess{a}{$\redeem$}{btc}
        \begin{call}
          {b}
          {look for signature in $\redeem$}
          {btc}{$\signature{redeem}{B}$}
        \end{call}
        \postlevel
        \begin{call}
          {b}
          {$\ecdsa.\rec(\signature{redeem}{B}, \encsig{redeem}{\Spend{a}{btc},B})$}
          {b}
          {$\s{a}$}
        \end{call}
        \mess{b}{redeem using $\s{a} + \s{b}$}{xmr}
      \end{sequencediagram}}
    \end{center}
  \end{mdframed}
  \caption{Happy path on-chain protocol.}
  \label{fig:onchain_seq}
\end{figure}

\subsection{Cross-chain DLEQ proof}\label{DLEQ}

The key generation phase depicted in \cref{fig:key_gen_protocol} shows both Alice and Bob producing a so-called \textit{cross-curve} DLEQ proof.
This construct is used to non-interactively prove in zero-knowledge that the public key pair $(\Spend{a}{btc}, \Spend{a}{xmr})$ has a common secret key $\s{a}$, and that the public key pair $(\Spend{b}{btc}, \Spend{b}{xmr})$ has a common secret key $\s{b}$.
Without these proofs, there would be no guarantee that the adaptor signatures $\encsig{redeem}{\Spend{a}{btc},B}$ and $\encsig{refund}{\Spend{b}{btc},A}$ which they later exchange actually commit to the expected signature-secret pairs.

Conventionally, this kind of proof can only be constructed for points on the same elliptic curve.
The idea of proving discrete logarithm equality across different groups comes from \cite{mrl-10}.
The algorithm proposed in \cite{mrl-10} is used in the original protocol by Gugger, but we elect to use something simpler based on the original idea so that it can be argued secure by the composition of sigma protocols\cite{berry}.

We built an experimental implementation of this proof\cite{x-curve-dleq-comit} to support our proof-of-concept implementation of this protocol\cite{xmr-btc-comit}.
We also contributed to a more general and efficient implementation of the same proof\cite{sigma-fun} which we intend to use in the future.

%% file: schemas/old_tranasction_schem.tex

\tikzstyle{line} = [draw, -latex']

\tikzstyle{lables} = [midway, above, sloped]


\tikzstyle{transaction} = 		[rectangle, text width=5em, text centered, rounded corners, minimum height=4.5em, minimum width=6em]

\tikzstyle{txLabel} = 			[rectangle, text width=5em, text centered, rounded corners, minimum height=2em, minimum width=6em]

\tikzstyle{IF} = 				[diamond, draw, fill=gray!20]

\tikzstyle{txSpendingLabel} = 	[rectangle, text width=5em, font=\scriptsize, text centered, rounded corners]

\newcommand\TxBox[5]{%
 	  	\node[rectangle] at ({#1},{#2}) (center) {};

   		\node[txLabel,above=1em of center] (nodeLabel) {#3};

   		\node[txSpendingLabel, below=1em of center] (spendingLabel) {#4};
   
	   	\node[{#5}, right=2.45em of spendingLabel] (spendingCondition){};
   
	   	\node[rectangle, draw, minimum height=1.5em, minimum width=6.3em, below=1em of center] (labelBox) {};
   
	   	\node[rectangle, draw, rounded corners, minimum height=5em, minimum width=7em, below=-0.em of center] (txBox) {};
}

\newcommand\TxBoxWithoutCondition[4]{%
 \TxBox{#1}{#2}{#3}{#4}{}
}

\newcommand\TxBoxWithCondition[4]{%
 \TxBox{#1}{#2}{#3}{#4}{IF}
}

\begin{figure}[htb]
	\centering
	\begin{tikzpicture}
		[-latex,     
    	on grid=true
    	]
		
		\begin{scope}[name prefix = txfund-]
			\TxBoxWithCondition{0}{-1.5}{$\lock{btc}$}{$a \land b$}
		\end{scope}
		
		\begin{scope}[name prefix = txredeem-]
			\TxBoxWithoutCondition{4}{0}{$\redeem$}{$\address{redeem}{A}$}
		\end{scope}
			
		\begin{scope}[name prefix = txcancel-]	
			\TxBoxWithCondition{4}{-3}{$\cancel$}{$a \land b$}
		\end{scope}		
		
		\begin{scope}[name prefix = txtake-]
			\TxBoxWithoutCondition{8}{-1.5}{$\refund$}{$\address{refund}{B}$}
		\end{scope}		
		
		\begin{scope}[name prefix = txpunish-]
			\TxBoxWithoutCondition{8}{-4}{$\punish$}{$\address{punish}{A}$}
		\end{scope}

		\draw [line] (txfund-spendingCondition.north) |- (txredeem-txBox.west) 
				node [labelBelow_small, font=\scriptsize]{$\A, \B$};
		
		\draw [line] (txfund-spendingCondition.south) |- (txcancel-txBox.west) 
		        node [labelAbove_small, font=\scriptsize] {$\checkRelative{t_1}$} 
				node [labelBelow_small,  font=\scriptsize]{$\A, \B$};
		
		\draw [line] (txcancel-spendingCondition.north) |- (txtake-txBox.west)
				node [labelBelow_small,  font=\scriptsize]{$\A, \B$};
						
		\draw [line] (txcancel-spendingCondition.south) |- (txpunish-txBox.west)
				node [labelAbove_small, font=\scriptsize] {$\checkRelative{t_2}$} 
				node [labelBelow_small,  font=\scriptsize]{$\A, \B$};
				
		\draw [-,dotted] (-1,-5) -- (9,-5);

		\begin{scope}[name prefix = xmr-lock-]
			\TxBoxWithCondition{0}{-6.5}{$\lock{xmr}$}{$\Spend{a}{} + \Spend{b}{}$}
		\end{scope}
		
		\begin{scope}[name prefix = xmr-redeem-]
			\TxBoxWithoutCondition{4}{-6.5}{$tx_{\textsf{redeem}}^{\textsf{xmr}}$}{$\textsf{Bob}$}
		\end{scope}
		
		\begin{scope}[name prefix = xmr-cancel-]
			\TxBoxWithoutCondition{4}{-8.5}{$tx_{\textsf{refund}}^{\textsf{xmr}}$}{$\textsf{Alice}$}
		\end{scope}
						
		\draw [dotted] (xmr-lock-spendingCondition.north) |- (xmr-redeem-txBox.west)
				node [labelBelow_small,  font=\scriptsize]{$\s{A}, \s{B}$};
												
		\draw [dotted] (xmr-lock-spendingCondition.south) |- (xmr-cancel-txBox.west)
				node [labelBelow_small,  font=\scriptsize]{$\s{A}, \s{B}$};
		
	\end{tikzpicture}
\caption{Transaction schema for BTC to XMR atomic swaps. \textit{Top}: Transaction schema for Bitcoin. \textit{Bottom}: Transaction schema for Monero. 
	\textit{Note: Monero view keys are omitted for clarity.}}
\label{fig:old_btc_protocol}
\end{figure}

%% file: new_protocol.tex
\section{XMR to BTC atomic swaps}\label{new_protocol}

In the section above we described an atomic swap protocol between Bitcoin and Monero. 

That protocol is appropriate for a use case in which the Service Provider (SP) is in the role of Alice, i.e. she offers buying XMR for BTC to her customers. 
Following the protocol as defined in \cref{old_protocol}, offering that kind of trade, allows the SP to lock up $\xmr$ (by publishing $\lock{xmr}$) only after the other party has locked up $\btc$ (by publishing $\lock{btc}$). 
This is safe for the SP as she knows that she will either be able to redeem (publish $\redeem$) or refund.

However, using that protocol to swap in the \textit{opposite} direction is not feasible i.e. an SP should not offer buying BTC for XMR.
The problem is that an SP (in the role of Bob) could be easily attacked: the taker (in the role of Bob) could agree on a trade with the SP, make him lock up funds on Bitcoin and then bail out at no cost. 
The SP could always refund his locked up BTC after some time, but he would have to pay for transaction fees to do so.
The taker's ability to make the SP incur in transaction fees without penalty would expose the SP to running out of funds over time, which is why we refer to this as a \textit{draining attack}.

We need a different protocol to allow an SP to offer BTC/XMR buy trades, since the original makes it a hard requirement for the party holding BTC to move first.
In the following sections we propose a new protocol which instead requires the party holding XMR to move first.
This depends on the development of adaptor signatures based on Monero's ring signature scheme, which is a work-in-progress and whose details are left out of the scope of this work.

\subsection{Protocol definition}

\input{schemas/xmr_schema.tex}

\input{schemas/btc_schema.tex}

\cref{fig:xmr_protocol} and \cref{fig:btc_protocol} show the transaction schema for Monero and Bitcoin respectively.
These diagrams are used to illustrate the relationships between different transactions on the same blockchain.
Transactions are represented as rectangles with rounded corners.
Transaction outputs are depicted as boxes inside transactions. 
The value of the output is written inside the output box and their spending conditions are written above and below arrows coming out of the output. 
For example, $x_A$ means the output holds $x$ coins owned by party $A$ and $(x_A \land x_B)$ means the output of amount $x$ is controlled by party $A$ and $B$.

With regard to the spending conditions, we define the following convention: the public keys of all required signatures are listed \textit{below} the arrow; other conditions, such as timelocks, appear \textit{above} the arrow.

\subsection{Creating Monero transactions}\label{xmr_create_transactions}

The transaction schema for Monero can be found in Figure \ref{fig:xmr_protocol}. 
Below we describe the naive 5-step protocol to create all transactions. 
An optimized implementation could reduce the number of steps by combining messages, but we refrain from doing so in the interest of clarity.  

\textbf{Step 1: } To construct the locking transaction \textit{$\txFundShortXmr{}$} the parties need to exchange some keys: 
Alice shares with Bob a public spend key $\xmrpk{A}$ and her private view key $\xmrvsk{A}$, as well as her funding source $tid_A$.
Bob shares with Alice his public spend key $\xmrpk{B}$ and his private view key $\xmrvsk{B}$.
They can now create \textit{$\txFundShortXmr{}$} locally with input $tid_A$ and an output with public spend key  $\xmrpk{A} + \xmrpk{B}$, private view key $\xmrvsk{A} + \xmrvsk{B}$.
Notably, Alice does not sign the transaction yet. 
Both parties now have a local copy of an unsigned \textit{$\txFundShortXmr{}$} which requires one signature from each party to spend its output.

\textbf{Step 2: } Both parties create the refund transaction \textit{$\txRefundShortXmr{}$} which spends from \textit{$\txFundShortXmr{}$} and returns the funds back to Alice.
Notably, they do not create the redeem transaction $\txTakeShortXmr{}$ in the same way, because they do not have to exchange any signatures on it. 
The key idea is that Bob will learn $\xmrsk{A}$ later on if Alice publishes $\txRedeemShortBtc{}$, allowing him to construct, sign and publish the redeem transaction $\txTakeShortXmr{}$ by himself.

\textbf{Step 3: }
Like in \cref{sec:btc-signing-phase} for the old protocol, adaptor signatures are used but this time on Bitcoin \textit{and} Monero. 
Alice generates a keypair $(\xmrrsk{A},\xmrrpk{A})$ and constructs a DLEQ proof for it for the same reasons presented in \cref{DLEQ}.
She sends $\xmrrpk{A}$ to Bob, which he uses as the encryption key to generate an adaptor signature on the refund transaction \textit{$\txRefundShortXmr{}$}.
Bob sends this adaptor signature to Alice.
If she were to ever publish \textit{$\txRefundShortXmr{}$} she would need to use this adaptor signature, leaking $\xmrrsk{A}$ to Bob.
This would allow him to execute an emergency refund on Bitcoin if Alice were misbehaving by attempting to take both the bitcoin and the monero.

\textbf{Steps 4+5: }Alice could now sign the locking transaction \textit{$\txFundShortXmr{}$} and publish it on the Monero blockchain with the assurance that she could get her funds back at any point by publishing \textit{$\txFundShortXmr{}$}. 
But these steps are not carried out until the two parties have collaborated on creating the Bitcoin transactions.

\subsection{Creating transactions for Bitcoin}\label{btc_create_transactions}

The transaction schema for Bitcoin can be found in Figure \ref{fig:btc_protocol}. 

\textbf{Step 1: } To prepare the locking transaction $\txFundShortBtc{}$, Alice shares a public key $pk_A$ with Bob. Bob shares his funding source $tid_B$ with Alice as well as a public key $pk_B$. 
Both parties can now create $\txFundShortBtc{}$ which spends from $tid_B$ into a multisignature output requiring two signatures: one for $pk_A$ and another one for $pk_B$.

\textbf{Step 2: } Knowing $\txFundShortBtc{}$, both parties can construct $\txRefundShortBtc{}$, a transaction which returns the bitcoin back to Bob after time $t_1$. 
They also construct $\txRedeemShortBtc{}$. 
This transaction sets the stage for Alice to be able to take the bitcoin.
It can be spent in two ways: (1) Alice can claim the coins after time $t_2$ by providing signatures for $pk_A$ and $pk_B$, and (2) Bob can still refund if he learns Alice's refund secret $\xmrrsk{A}$ and uses it with his own public key $pk_B$. 
Bob would learn $\xmrrsk{A}$ if Alice publishes $\txRefundShortBtc{}$, using the adaptor signature generated in step 3 of the Monero transaction creation protocol above.

\textbf{Step 3: } Having constructed $\txRedeemShortBtc{}$, both parties can create $\txTakeShortBtc{}$, which spends from it and can be published after time $t_2$ giving the funds to Alice.

\textbf{Step 4: } For safety purposes, transactions are signed in reverse order of publication.
To that end, Alice and Bob collaboratively sign $\txTakeShortBtc{}$.
Only Bob sends his signature to Alice because she is the one that would care to publish this transaction, since it benefits her.
There is no need to create $\txEmergencyRefundShortBtc$ which would require a signature from $\xmrrpk{A}$ and $pk_B$. Bob will be able to create and sign this transaction by himself if the situation allows.

\textbf{Step 5: } Alice and Bob sign $\txRefundShortBtc$ collaboratively.
Only Alice shares her signature with Bob because he is the only one interested in ever being able to take his bitcoin back.
Bob also generates an adaptor signature on $\txRedeemShortBtc$ for his public key $pk_B$ encrypted under $\xmrpk{A}$ and sends it to Alice.
This adaptor signature ensures the atomicity of the swap: if Alice publishes $\txRedeemShortBtc$ she will need to decrypt and use the adaptor signature, leaking $\xmrsk{A}$, which he would use to take the monero.

\textbf{Step 6+7: } Bob is now ready to sign and publish $\txFundShortBtc$.
He still must wait for Alice to lock her monero first by publishing $\txFundShortXmr$, finishing steps 4 and 5 of \cref{xmr_create_transactions}.
Once Alice has committed her funds to the Monero blockchain, Bob is safe to do the same on Bitcoin.

\subsection{Protocol execution}
The content of this section is still work-in-progress. Hence we do not delve deeper into the cryptography which is needed to create adaptor signatures on Monero. 
Instead, we continue describing things on a high level.

\subsubsection{Scenario}
The motivation behind this protocol is to allow the party holding XMR and wanting BTC to move first. 
In this scenario, Alice holds XMR and wants to receive BTC. Conversely, Bob holds BTC and wants to receive XMR.

After successfully building and signing transactions following the steps outlined in \cref{xmr_create_transactions} and \cref{btc_create_transactions}, Alice and Bob are ready to go on-chain.

\subsubsection{Happy path}

Alice publishes her locking transaction $\txFundShortXmr$ knowing that she can always receive her funds back by cancelling the swap and publishing $\txRefundShortXmr$. 
Once Bob is happy with the amount of confirmations on $\txFundShortXmr$ he follows suit and publishes the locking transaction on Bitcoin $\txFundShortBtc$. 
Given sufficient confirmations on $\txFundShortBtc$ and enough time until $t_1$, Alice publishes the redeem transaction $\txRedeemShortBtc$. In doing so, she leaks $\xmrsk{A}$ to Bob.
Alice cannot immediately claim the bitcoin for herself but has to wait until time $t_2$. 
In the meantime, Bob has until time $t_2$ to safely take the monero by using $\xmrsk{A}$ to create and sign $\txTakeShortXmr$, and publishing it on the blockchain. 
Once time $t_2$ is reached. Alice can finally take the bitcoin by publishing $\txTakeShortBtc$, completing the atomic swap. 

\subsubsection{One party is unresponsive}

At any point in time during the execution phase either party could become inactive. 
In order to prevent money loss, both parties have mechanisms at their disposal to refund. 

For instance, Alice could publish her locking transaction $\txFundShortXmr$ and then see that Bob never moves forward with the publication of his locking transaction $\txFundShortBtc$. 
As depicted in \cref{fig:xmr_protocol}, $\txRefundShortXmr$ requires signatures on $\xmrpk{A}$ and $\xmrpk{B}$.
Alice can use her own secret key $\xmrsk{A}$ to produce one of the signatures, and decrypt Bob's adaptor signature using $\xmrrsk{A}$ to produce the other.
She would then publish $\txRefundShortXmr$, taking back her monero.

Similarly, if Bob does publish $\txFundShortBtc$, but Alice fails to continue by publishing $\txRedeemShortBtc$ before time $t_1$, Bob can then take his bitcoin by publishing $\txRefundShortBtc$, since he either has or can produce or the signatures needed for it to be valid.

\subsubsection{Alice tries to cheat}

There exists an edge case in which Alice can attempt to take both assets. 
This is possible after both parties have published their respective locking transactions $\txFundShortXmr$ and $\txFundShortBtc$. 
Alice may attempt to redeem the bitcoin by publishing $\txRedeemShortBtc$ \textit{and} refund the monero by publishing $\txRefundShortXmr$. 
Fortunately, the publication of $\txRefundShortXmr$ would leak $\xmrsk{A}$ to Bob which would allow him to create, sign and publish $\txEmergencyRefundShortBtc$ to execute an emergency refund, at least until time $t_2$.
The result would be equivalent to having executed a \textit{normal} refund. 
Bob therefore remains protected, but this possibility imposes a strong requirement for him to stay online at all times.

%% file: schemas/xmr_schema.tex
\newcommand{\txFundShortXmr}{\mathtt{XMR_{l}}}
\newcommand{\txRefundShortXmr}{\mathtt{XMR_{c}}}
\newcommand{\txTakeShortXmr}{\mathtt{XMR_{r}}}
\newcommand{\SA}{\mathtt{S_A}}
\newcommand{\VA}{\mathtt{v_A}}
\newcommand{\SB}{\mathtt{S_B}}
\newcommand{\VB}{\mathtt{v_B}}

\newcommand{\xmrpk}[1]{S_{#1}}
\newcommand{\xmrsk}[1]{s_{#1}}

\newcommand{\xmrvpk}[1]{V_{#1}}
\newcommand{\xmrvsk}[1]{v_{#1}}

\newcommand{\xmrrpk}[1]{R_{#1}}
\newcommand{\xmrrsk}[1]{r_{#1}}

\newcommand{\sign}{\mathsf{sign}}

\begin{figure}[htbp]
	\centering
	\begin{tikzpicture}[auto]
	
	\begin{scope}[on background layer]
		\node (TxFund) {$\txFundShortXmr$};
		\node [node distance=0.7cm, below of=TxFund, font=\scriptsize](txFundCondition){$x_A$};
		
		\node[IF,node distance=0.77cm, right of=txFundCondition](SplitIF){};
		\node[output,inner sep=1,outer sep=0, fit = (txFundCondition)(SplitIF)](Splitout){};
		\node [TXsmall, fit=(TxFund)(Splitout)](FundingBox){};
	
	\end{scope}
	
	
	\begin{scope}[on background layer]
		\node[node distance=3.4cm,right of=FundingBox](TxTakeBox){};
		\node[node distance=0.35cm,above of=TxTakeBox, align=center](TxTake){$\txTakeShortXmr{}$};

		\node[node distance =0.14cm, below of=TxTakeBox, align=left, font=\scriptsize](TxTakeCondition){$x_B$};
	
		\node[node distance=0.1cm, right of=TxTakeCondition](TxTakeLocation){};		
		\node[output,inner sep=1,outer sep=1, fit = (TxTakeCondition)(TxTakeLocation)](TxTakeSplitout){};
		
		\node[TXsmall, fit=(TxTake)(TxTakeSplitout)](TxTakeSplitbox){};
	\end{scope}

	
	\begin{scope}[on background layer]
		\node[node distance=1.8cm,below of=TxTakeBox](TxRefundBox){};
		\node[node distance=0.35cm,above of=TxRefundBox, align=center](TxRefund){$\txRefundShortXmr{}$};
		\node[node distance =0.14cm, below of=TxRefundBox, align=left, font=\scriptsize](TxRefundCondition){$x_A$};
		\node[node distance=0.1cm, right of=TxRefundCondition](TxRefundLocation){};
		\node[output,inner sep=1,outer sep=0, fit=(TxRefundCondition)(TxRefundLocation)](TxRefundSplitout){};
		\node[TXsmall, fit=(TxRefund)(TxRefundSplitout)](TxRefundSplitbox){};
	\end{scope}
	
	\path [dotted,line] (SplitIF) |- node [labelBelow_small, font=\scriptsize] {$\SA, \SB$}(TxTakeSplitbox);
	\path [line] (SplitIF) |- node [labelBelow_small, font=\scriptsize] {$\SA, \SB$}(TxRefundSplitbox);

	
	\node [node distance =4.1cm, below of=TxFund](prepareFund){
		\begin{minipage}{3cm}
		\begin{center}
			\small{\textbf{1. Create} $[\txFundShortXmr{}]$}\\~\\
				$\xrightarrow[ ]{\SA,\VA} $\\
				$\xleftarrow[ ]{\SB,\VB} $ 
		\end{center}
		\end{minipage}
	};
	
	\node [node distance =2cm, below of=TxRefundSplitbox](prepareCommmit){
		\begin{minipage}{3cm}
		\begin{center}
			\small{\textbf{2. Create} $[\txRefundShortXmr{}]$}\\~\\
			No communication. \\
			$ $
		\end{center}
	\end{minipage}
	};

	\node [node distance =2.5cm, below of=prepareFund](signFund){
		\begin{minipage}{3cm}
		\begin{center}
			\small{\textbf{4. Sign } $[\txFundShortXmr{}]$}\\~\\
			Alice $\sig(\xmrsk{A}, [\txFundShortXmr{}]) $\\
			$ $ \\
			~\\
			\small{\textbf{5. Publish }$[\txFundShortXmr{}]$}
		\end{center}	
		\end{minipage}
	};
	
	\node [node distance =2.5cm, below of=prepareCommmit](signCommit){
		\begin{minipage}{3cm}
		\begin{center}
				\small{\textbf{3. Sign } $[\txRefundShortXmr{}]$}\\~\\
				$\xrightarrow[ ]{\xmrrpk{A}} $ \\
				$\xleftarrow[ ]{\enc\sig(\xmrsk{B}, \xmrrpk{A}, [\txRefundShortXmr{}])} $ \\~\\~\\
		\end{center}
	\end{minipage}	
	};
	
	\node[minimum height= 8.5cm,fit=(FundingBox) (prepareFund) (signFund), fill=gray!40, rounded corners=1mm, fill opacity=.2]{};
	\node[minimum height= 8.5cm,fit=(TxTakeSplitbox) (prepareCommmit) (signCommit), fill=gray!40, rounded corners=1mm, fill opacity=.2]{};

	\end{tikzpicture}
\caption{Monero transaction schema.}
\label{fig:xmr_protocol}
\end{figure}
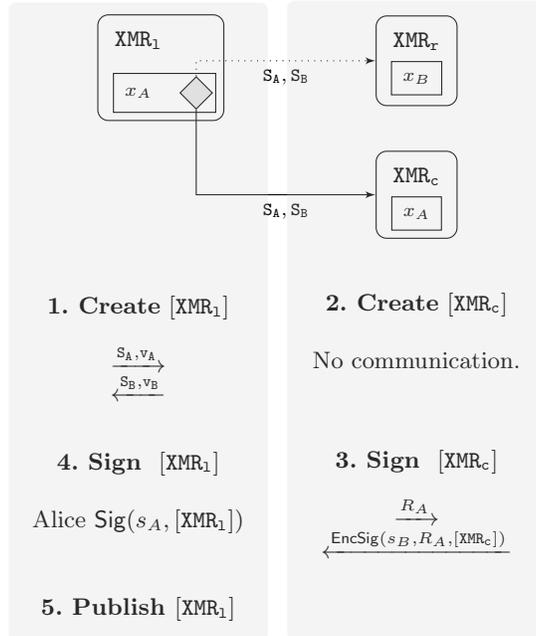

%% file: schemas/btc_schema.tex
\newcommand{\txFundShortBtc}{\mathtt{BTC_l}}
\newcommand{\txRedeemShortBtc}{\mathtt{BTC_r}}
\newcommand{\txTakeShortBtc}{\mathtt{BTC_t}}
\newcommand{\txEmergencyRefundShortBtc}{\mathtt{BTC_e}}
\newcommand{\txRefundShortBtc}{\mathtt{BTC_c}}

\newcommand{\xa}{\mathtt{x_A}}
\newcommand{\xb}{\mathtt{x_B}}
\newcommand{\aandb}{\mathtt{(\xa \land \xb)}}

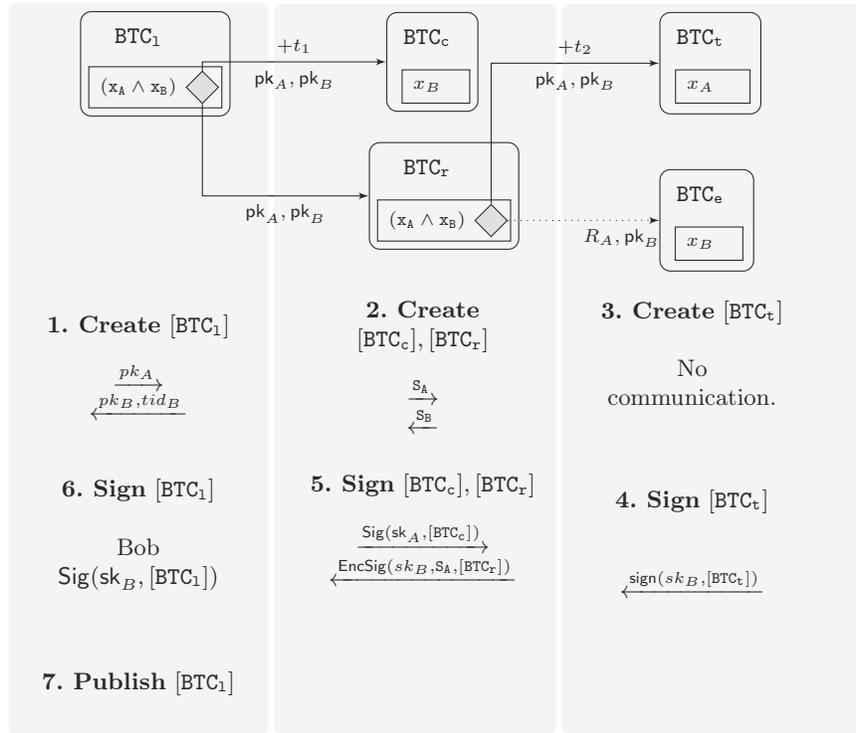
\begin{figure}[htb]
	\centering
	\begin{tikzpicture}[auto, node distance=2 cm]
	

	\node (TxFund) {$\txFundShortBtc$};
	\node [node distance=0.2cm, below=of TxFund, font=\scriptsize](txFundCondition){$\aandb$};

	\begin{scope}[on background layer]
		
		\node[IF,node distance=0cm, right=of txFundCondition](SplitIF){};
		\node[output,inner sep=1,outer sep=0, fit = (txFundCondition)(SplitIF)](Splitout){};
		\node [TXsmall, fit=(TxFund)(Splitout)](FundingBox){};
	
	\end{scope}
	
	
	\begin{scope}[on background layer]
		\node[node distance=2.5cm,right=of FundingBox](TxRefundBox){};
		\node[node distance=0cm,above=of TxRefundBox, align=center](TxRefund){$\txRefundShortBtc{}$};
		\node[node distance =0cm, below=of TxRefundBox, align=left, font=\scriptsize](TxRefundCondition){$x_B$};
		\node[node distance=0cm, right=of TxRefundCondition](TxRefundLocation){};
		\node[output,inner sep=1,outer sep=0, fit=(TxRefundCondition)(TxRefundLocation)](TxRefundSplitout){};
		\node[TXsmall, fit=(TxRefund)(TxRefundSplitout)](TxRefundSplitbox){};
	\end{scope}
	
	
	\begin{scope}[on background layer]
		\node[node distance=1.5cm,below=of TxRefundBox](TxRedeemBox){};
		\node[node distance=0cm,above=of TxRedeemBox, align=center](TxRedeem){$\txRedeemShortBtc{}$};

		\node[node distance =0cm, below=of TxRedeemBox, align=center, font=\scriptsize](TxRedeemCondition){$\aandb$};

		\node[IF,node distance=0cm, right=of TxRedeemCondition](TxRedeemSplitIF){};
		\node[output,inner sep=1,outer sep=0, fit = (TxRedeemCondition)(TxRedeemSplitIF)](TxRedeemSplitout){};
		\node [TXsmall, fit=(TxRedeem)(TxRedeemSplitout)](RedeemBox){};
		
	\end{scope}

	
	\begin{scope}[on background layer]
		\node[node distance=3.4cm,right=of TxRefundBox](TxTakeBox){};
		\node[node distance=0cm,above=of TxTakeBox, align=center](TxTake){$\txTakeShortBtc{}$};

		\node[node distance=0cm, below=of TxTakeBox, align=center, font=\scriptsize](TxTakeCondition){$x_A$};
	
		\node[node distance=0.1cm, right=of TxTakeCondition](TxTakeLocation){};		
		\node[output,inner sep=1,outer sep=1, fit = (TxTakeCondition)(TxTakeLocation)](TxTakeSplitout){};
		
		\node[TXsmall, fit=(TxTake)(TxTakeSplitout)](TxTakeSplitbox){};
	\end{scope}
	
	
	\begin{scope}[on background layer]
		\node[node distance=1.86cm,below=of TxTakeBox](TxERBox){};
		\node[node distance=0cm,above=of TxERBox, align=center](TxER){$\txEmergencyRefundShortBtc{}$};

		\node[node distance=0cm, below=of TxERBox, align=center, font=\scriptsize](TxERCondition){$x_B$};
	
		\node[node distance=0cm, right=of TxERCondition](TxERLocation){};		
		\node[output,inner sep=1,outer sep=1, fit = (TxERCondition)(TxERLocation)](TxERSplitout){};
		
		\node[TXsmall, fit=(TxER)(TxERSplitout)](TxERSplitbox){};
	\end{scope}

	\path [line] (SplitIF) |- (TxRefundSplitbox) node [labelAbove_small, font=\scriptsize] {$\checkRelative{t_1}$} node [labelBelow_small,  font=\scriptsize]{$\pk_A, \pk_B$};

	\path [line] (SplitIF) |- node [labelBelow_small, font=\scriptsize] {$\pk_A, \pk_B$}(RedeemBox);
	
	\path [line] (TxRedeemSplitIF) |- (TxTakeSplitbox) node [labelAbove_small, font=\scriptsize] {$\checkRelative{t_2}$} node [labelBelow_small,  font=\scriptsize]{$\pk_A, \pk_B$};
	
	\path [dotted,line] (TxRedeemSplitIF) -- node [labelBelow_small, font=\scriptsize] {$\xmrrpk{A},\pk_B$}(TxERSplitbox);
	
	
	\node [node distance =3.35cm, below=of TxFund](createFund){
		\begin{minipage}{2.5cm}
		\begin{center}
			\small{\textbf{1. Create} $[\txFundShortBtc{}]$}\\~\\
				$\xrightarrow[ ]{pk_A} $\\
				$\xleftarrow[ ]{pk_B, tid_B} $ 
		\end{center}
		\end{minipage}
	};

	\node [node distance=1.2cm, right=of createFund](createCancel){
		\begin{minipage}{\minipagesize}
		\begin{center}
			\small{\textbf{2. Create} $[\txRefundShortBtc{}], [\txRedeemShortBtc{}]$}\\~\\
				$\xrightarrow[ ]{\SA} $\\
				$\xleftarrow[ ]{\SB} $ 
		\end{center}
		\end{minipage}
	};

	\node [node distance=1cm, right=of createCancel](createTake){

		\begin{minipage}{2.5cm}
		\begin{center}
			\small{\textbf{3. Create} $[\txTakeShortBtc{}]$}\\~\\
				No communication.\\
				$ $ 
		\end{center}
		\end{minipage}
		
	};
	
	\node [node distance =0.5cm, below=of createTake](signTake){
		\begin{minipage}{2.5cm}
		\begin{center}
			\small{\textbf{4. Sign} $[\txTakeShortBtc{}]$}\\~\\
				$ $ \\ 
				$\xleftarrow[ ]{\sign(sk_B, [\txTakeShortBtc{}])} $ 
		\end{center}
		\end{minipage}
	};
	
	\node [node distance =0.3cm, below=of createCancel](signRedeem){
		\begin{minipage}{3.1cm}
		\begin{center}
			\small{\textbf{5. Sign} $[\txRefundShortBtc{}],[\txRedeemShortBtc{}]$}\\~\\
				$\xrightarrow[ ]{\sig(\sk_A, [\txRefundShortBtc{}])} $\\
				$\xleftarrow[ ]{\enc\sig(sk_B, \SA, [\txRedeemShortBtc{}])} $ 
		\end{center}
		\end{minipage}
	};
	
	\node [node distance =0.55cm, below=of createFund](signFund){
		\begin{minipage}{\minipagesize}
		\begin{center}
			\small{\textbf{6. Sign} $[\txFundShortBtc{}]$}\\~\\
				Bob $\sig(\sk_B, [\txFundShortBtc])$ \\
				$ $ 
		\end{center}
		\end{minipage}
	};
	
	\node [node distance =0.55cm, below=of signFund](publish){
		\begin{minipage}{3cm}
		\begin{center}
			\small{\textbf{7. Publish} $[\txFundShortBtc{}]$}\\~\\

		\end{center}
		\end{minipage}
	};
	
	\node [node distance =0.3cm, right=of publish](space2ndcol){
		\begin{minipage}{0.4cm}
		\vspace{0.2cm}
		\begin{center}
			$\,$  \\~\\
		\end{center}
		\end{minipage}
	};

	\node [node distance =3.56cm, right=of space2ndcol](space3rdCol){
		\begin{minipage}{3.cm}
		\vspace{0.2cm}
		\begin{center}
			$\,$  \\~\\
		\end{center}
		\end{minipage}
	};

	\node[minimum height=8.5cm,
	fit=(FundingBox) (prepareFund) (signFund)(publish), fill=gray!40, rounded corners=1mm, fill opacity=.2]{};
	
	\node[minimum height=8.5cm,
	fit=(TxRefundSplitbox)(RedeemBox) (signRedeem)(space2ndcol), fill=gray!40, rounded corners=1mm, fill opacity=.2]{};

	\node[minimum height= 8.5cm,minimum width=\textwidth/3,
	fit=(TxTakeSplitbox)(TxERSplitbox)(signTake)
	(space3rdCol)
	, fill=gray!40, rounded corners=1mm, fill opacity=.2]{};


	\end{tikzpicture}
\caption{Bitcoin transaction schema.}
\label{fig:btc_protocol}
\end{figure}

%% file: conclusion.tex
\section{Conclusion}

Atomic swaps constitute the main mechanism to bridge the gap between unrelated blockchains without violating the core principles of censorship resistance, permissionlessness and pseudonymity originally championed by Bitcoin.
Up until recently, their application was believed to be exclusive to blockchains with very particular characteristics.
Advances in cryptography have lowered the barrier to entry, allowing for new protocols to be devised in order to connect blockchains that were originally thought to be incompatible. One such example is the \textit{Bitcoin–Monero Cross-chain Atomic Swap} by Gugger\cite{gugger2020}, which has inspired the development of applications such as \cite{farcaster} and \cite{xmr-btc-comit}. 

In this work, we give a high-level sketch of a new protocol which expands on the ideas of the original to serve a new use case.
In particular, by applying adaptor signatures to the Monero signature scheme, we make possible atomic swaps in which the party holding BTC is no longer the one vulnerable to draining attacks.
A real-world service provider could therefore leverage both protocols to put up buy and sell BTC/XMR offers as a market maker.

This proposal hinges on the viability of using adaptor signatures on Monero, a topic which we do not discuss here, but one which is being researched at the time of writing.